%% file: Li.ACC.2023.tex
\documentclass[letterpaper, 10 pt, conference]{ieeeconf}

\IEEEoverridecommandlockouts           

\usepackage{setspace}

\usepackage{graphicx}
\usepackage{lipsum}
\usepackage{caption}
\usepackage{subcaption}
\usepackage{sidecap}
\usepackage{amsmath}
\usepackage{amsfonts}
\usepackage{bm}
\usepackage{xfrac}

\PassOptionsToPackage{hyphens}{url}
\usepackage{url}
\usepackage{hyperref}
\hypersetup{breaklinks=true}

\usepackage{xcolor}
\usepackage{float}
\usepackage{algorithm}
\usepackage{algpseudocode}
\usepackage{wrapfig}
\usepackage{makecell}
\usepackage{tabularray}

\graphicspath{ {./figure/} }

\title{\LARGE \bf
Differentially Private Timeseries Forecasts for Networked Control
}

\author{Po-han Li$^{1}$, Sandeep P. Chinchali$^{1}$, Ufuk Topcu$^{2}$
\thanks{$^{1}$Department of Electrical  and Computer Engineering, The University of Texas at Austin, $^{2}$Oden Institute for Computational Engineering and Sciences, The University of Texas at Austin,
        {\tt\small \{pohanli, sandeepc, utopcu\}@utexas.edu}}}

\begin{document}

\input{section/Preamble}

\maketitle

\begin{abstract}
\input{section/Abstract}
\end{abstract}

\section{Introduction}
\input{section/Introduction}

\section{Preliminaries}
\input{section/Preliminaries}

\section{System Model}
\input{section/SystemModel}

\section{Incentive Allocation for Differentially Private Forecasts}
\input{section/Method}

\section{Experiments}
\input{section/Experiments}

\section{Conclusions and Future Work}
\input{section/Conclusion}

\section*{Acknowledgement}
\input{section/Acknowledgement}

\bibliographystyle{ieeetr}
\bibliography{bibtex/swarm, bibtex/external}


\end{document}

%% file: section/Preamble.tex
\newcommand{\sandeep}[1] {{\color{red} \textbf{[Sandeep]: #1}}}
\newcommand{\pohan}[1] {{\color{blue} \textbf{#1}}}
\newcommand{\ufuk}[1]{{\color{green} \textbf{[Ufuk]: #1}}}

\newcommand{\xtdim}{n}
\newcommand{\utdim}{m}
\newcommand{\stdim}{p}
\newcommand{\horizon}{T}
\newcommand{\currenttime}{t}
\newcommand{\srcnum}{n_{\mathrm{src}}}
\newcommand{\src}{i_{\mathrm{src}}}
\newcommand{\Sgt}[1]{S_{#1}}
\newcommand{\Shat}[1]{\hat{S}_{#1}}

\newcommand{\action}[1]{u_{#1}}
\newcommand{\actionhat}[1]{\hat{u}_{#1}}
\newcommand{\state}[1]{x_{#1}}
\newcommand{\ctrlcost}{J^c}
\newcommand{\A}{\bm{A}}
\newcommand{\B}{\bm{B}}
\newcommand{\C}{\bm{C}}
\newcommand{\Q}{\bm{Q}}
\newcommand{\R}{\bm{R}}
\newcommand{\M}{\bm{M}}
\newcommand{\N}{\bm{N}}
\newcommand{\K}{\bm{K}}
\newcommand{\LL}{\bm{L}}
\newcommand{\Lone}{\bm{L}_1}
\newcommand{\Ltwo}{\bm{L}_2}
\newcommand{\codesignPsi}{\bm{\Psi}}
\newcommand{\blockdiag}{\mathrm{BlockDiag}}

\newcommand{\deltaS}[1]{\Delta\hat{S}_{#1}}
\newcommand{\deltaSDP}[1]{\deltaS{#1}^\mathrm{dp}}
\newcommand{\deltaSForecast}[1]{\deltaS{#1}^\mathrm{pc}}
\newcommand{\Laplace}[2]{\mathrm{Lap}\left(#1, #2\right)}
\newcommand{\Normal}[2]{\mathcal{N}(#1, #2)}
\newcommand{\Ssen}[1]{\mathrm{Sen}(S_{#1})}
\newcommand{\eprivacy}[1]{\epsilon_{#1}}
\newcommand{\incentive}[1]{\rho_{#1}}
\newcommand{\maxeprivacy}[1]{\alpha_{#1}}
\newcommand{\midincent}[1]{\gamma_{#1}}
\newcommand{\edecay}[1]{\beta_{#1}}
\newcommand{\indentity}[1]{\mathbb{I}_{#1}}
\newcommand{\LaplaceVar}[1]{\sigma^2( \incentive{#1} )}
\newcommand{\coeff}[1]{\bm{c_{#1}}}
\newcommand{\window}[1]{w_{#1}}

\newcommand{\deltanoise}{\delta}
\newcommand{\domevec}{v_1}
\newcommand{\domeval}{\lambda_1}

\newcommand{\gencostfunc}{J^c}
\newcommand{\ineqfunc}[1]{f_{#1}}
\newcommand{\eqfunc}[1]{g_{#1}}
\newcommand{\ninequal}{n_\mathrm{ineq}}
\newcommand{\nequal}{n_\mathrm{eq}}

\newcommand{\advfunc}{h_\mathrm{adv}}
\newcommand{\unit}{\mathrm{Unit}}

\newcommand{\acs}{\textit{ACS}}
\newcommand{\uniform}{\textit{Uniform}}

%% file: section/Abstract.tex
We analyze a cost-minimization problem in which the controller relies on an imperfect timeseries forecast. 
Forecasting models generate imperfect forecasts because they use anonymization noise to protect input data privacy. However, this noise increases the control cost.
We consider a scenario where the controller pays forecasting models incentives to reduce the noise and combines the forecasts into one. The controller then uses the forecast to make control decisions.
Thus, forecasting models face a trade-off between accepting incentives and protecting privacy. 
We propose an approach to allocate economic incentives and minimize costs. 
We solve a biconvex optimization problem on linear quadratic regulators and compare our approach to a uniform incentive allocation scheme.
The resulting solution reduces control costs by 2.5 and 2.7 times for the synthetic timeseries and the Uber demand forecast, respectively.



%% file: section/Introduction.tex
Controllers relying on timeseries forecasts have applications in power grid operations \cite{cheng2021data, NIPS2017Donti}, cellular network traffic scheduling \cite{chinchali2018cellular}, and taxi fleet routing. 
For example, power grid operators use electricity demand forecasts to charge the batteries, and cellular network providers use city-wide mobility forecasts to allocate bandwidth among base stations. Ridesharing companies use customer demand forecasts to assign taxis to different queues in a city, which is the example we used in our experiments. 
In these examples, the controller relies on an accurate future timeseries forecasts to make control decisions and then minimizes its control cost. 
The control cost is a function of the forecasting error, which is the difference between the forecast and the actual timeseries. 

We use the system model in Fig. \ref{fig:systemgraph} to characterize systems with a controller relying on a timeseries forecast. Multiple forecasting models transmit timeseries forecasts to the controller through a network. The controller can only receive forecasts from the forecasting models but cannot influence the timeseries.
Fig. \ref{fig:systemgraph} shows the system model of the controller and forecasting models. 
The forecasts have different intrinsic errors, and the controller combines the forecasts to generate a more accurate one used for control decisions. 
However, forecasting models tend to add anonymization noise to the forecasts to protect data privacy, thus resulting in more inaccurate forecasts and additional control costs. 
We analyze a scenario where the controller is able to pay the forecasting models an economic incentive, such as money, to reduce their noise. 
The more incentive paid, the more accurate the forecasts, and the lower the control costs. 
The problem of trade-offs between accepting incentives or protecting privacy for forecasting models then arises, and different models have different tendencies to choose between the two. 
Given the trade-off, a question arises: how a controller incentivizes multiple forecasting models to reduce the noise and thereby minimizes the resulting cost while the models preserve sufficient privacy? 
We study the cost minimization problem with multiple forecasts using the linear quadratic regulator (LQR) as the controller and differential privacy as the privacy preserving mechanism. 
We use \textit{prediction errors} to refer to intrinsic errors due to the unpredictable future, and \textit{forecasting errors} refers to the overall error due to prediction and differential privacy mechanisms.

\input{figure_latex/system_graph}

\subsection{Related Work}
Previous work analyzes controllers, referred to as \textit{input-driven controllers}, which rely on accurate future timeseries forecasts \cite{cheng2021data, li2022adversarial}.
The authors study how to compress timeseries for input-driven controllers \cite{cheng2021data} and show that adversarial perturbations dramatically increase the control costs \cite{li2022adversarial}. 
The key difference between these works and ours is that their settings contain a \textit{single} forecast source, while ours focuses on reducing differentially private noise between \textit{multiple} sources. 

The differential privacy mechanism preserves the privacy of the input data of a function by adding random noise to the output \cite{DifferentialPrivacyAutomata}. 
The noise resulting from the differential privacy mechanism can perturb the estimation of control-relevant states and thus increase the control cost. 
Previous work studies the trade-off between data privacy and performance of control systems \cite{DP_linear_distributed, DPconsensus, cheng2022task, hassan2019differential}. 
Previous work focuses on the performance of control systems \cite{DP_linear_distributed}, consensus algorithms \cite{DPconsensus}, and cyber-physical systems \cite{hassan2019differential}. 
Another work studies controllers with multiple sensors \cite{ICRA_DP}.
Our work also analyzes the trade-off between data privacy and performance among multiple sources of signals, but our controller interacts with sources using incentives. Thus, it results in another trade-off of forecasting models to receive incentives or protect privacy. 

Other literature also studies settings that encourage less privacy noise \cite{PrivacyDrivenTruthful, MobileCrowdSensing, EHealthcareIoT, LargeGames}. 
However, our work focuses on the setting of a networked controller that pays incentives to forecasting models, while they focus on mobile crowdsensing systems \cite{PrivacyDrivenTruthful, MobileCrowdSensing}, e-healthcare devices \cite{EHealthcareIoT}, and recommender mechanisms \cite{LargeGames}. 

\subsection{Insights and Contributions}
Lee et al. study the appropriate range of differential privacy noise levels in a variety of applications \cite{HowMuchIsEnough}, and their work gives us the insight that the noise levels do not need to be \textit{a priori} fixed. 
In fact, forecasting models can choose their own noise levels from a range. 
Thus, the noise levels can be reduced by an economic incentive from the controller since it is a flexible choice. 
The more incentive the controller pays, the less noise, resulting in a more accurate forecast. 
The proper use of incentives to minimize costs is an emerging issue. 

Based on the insights, our key contributions are three-fold.
First, we characterize the trade-off between privacy and incentives. 
Second, we formulate an incentive allocation problem, in which the controller optimizes the allocation of incentives and combines the forecasts to minimize its control cost. 
Lastly, we use the linear quadratic regulator as an example of this scenario and solve a biconvex optimization problem with guaranteed local optimality. We then numerically show that our method can significantly reduce the control cost on synthetic ARIMA timeseries and the Uber demand forecast.

%% file: figure_latex/system_graph.tex
\begin{figure}[t]
\centering
\includegraphics[width=0.5\textwidth]{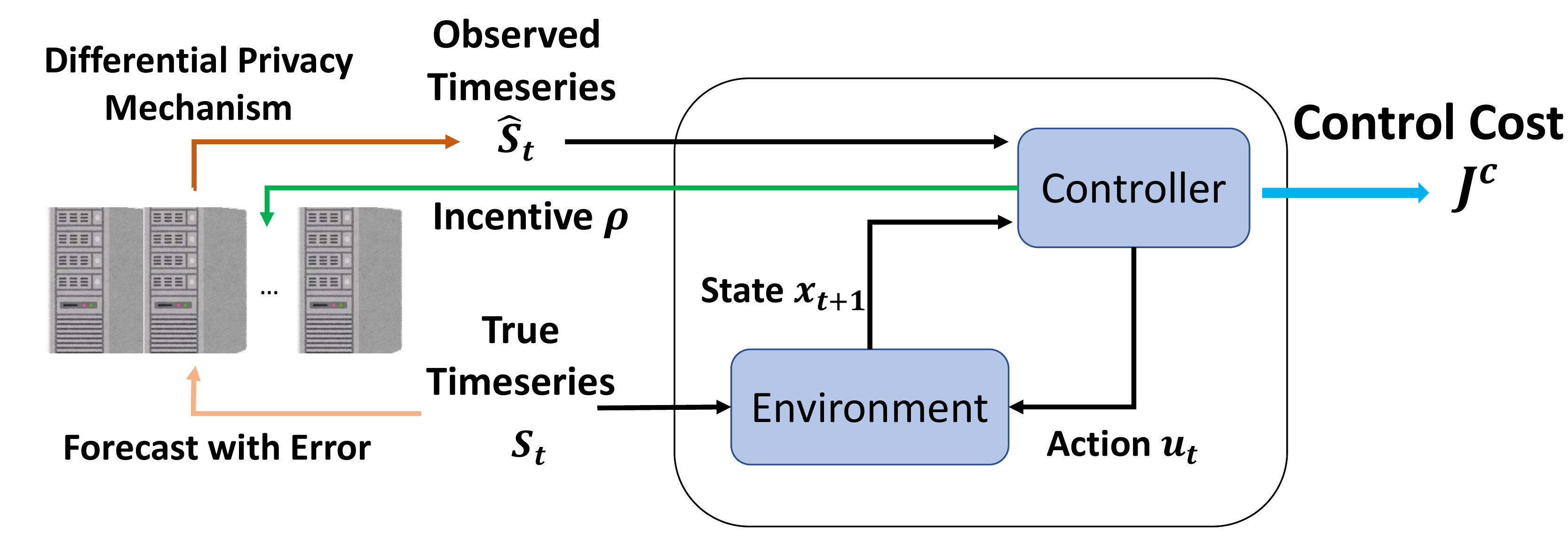}
\caption{\small
    \textbf{Differentially private timeseries forecasts for networked control.}
    A controller incentivizes multiple forecasting models to reduce differentially private noise and combines the resulting forecasts to minimize its control cost. 
    At time $\currenttime$, the controller pays incentives $\incentive{}$ to forecasters to reduce the noise caused by differential privacy mechanisms. 
    It then observes its control state $\state{\currenttime}$ and multiple timeseries forecasts $\Shat{\currenttime}$, which include prediction errors and differentially private noise. The controller takes action $\action{\currenttime}$ to minimize the cost $\ctrlcost$. The dynamics of next state $\state{\currenttime+1}$ are determined by previous state $\state{\currenttime}$, action $\action{\currenttime}$, and the true future timeseries $\Sgt{\currenttime}$. 
}
\label{fig:systemgraph}
\vspace{-2em}
\end{figure}

%% file: section/Preliminaries.tex
\subsection{$\epsilon$-Local Differential Privacy}
Let $\epsilon$ be a positive real number and $\mathcal{A}$ be a forecasting model with randomness in its output \cite{PrivacyAtScaleLDP}. 
$\mathcal{A}$ takes a model's private data as input and maps it to the range $\mathrm{im}\mathcal{A}$. 
The forecasting model $\mathcal{A}$ is said to provide $\epsilon$-local differential privacy if for all pairs of private input data $x$ and $x' \in \mathbb{X}$ and all possible subsets $S$ of $\mathrm{im}\mathcal{A}$:
\begin{equation}
Pr\{\mathcal{A}(x)\in S\} \leq e^{\epsilon} \times Pr\{\mathcal{A}(x')\in S\}, 
\end{equation}
where the probability is due to the randomness of the forecasting model $\mathcal{A}$, and $\mathbb{X}$ is the universal set of all inputs. 
The main difference between standard (global) differential privacy and local differential privacy is that the former takes all models' private data pairs, and the latter takes a single model's private data pairs. 
Here, $\epsilon$ is called the privacy budget. The larger the privacy budget $\epsilon$, the less privacy preserved.

\subsection{Laplace Mechanism}
The Laplace mechanism is commonly used for obtaining $\eprivacy{}$-local differential privacy. It perturbs the output of functions with random noise to protect the original information of the input. We use the definition from \cite{SensitivityinPrivate, AlgorithmicDP}. Given any arbitrary function $A: \mathbb{X} \rightarrow \mathbb{R}^d$, the sensitivity of $A$ under the $L_1$ norm is defined as: 
$$\mathrm{Sen}(A)=\max_{x,x'\in  \mathbb{X}} \|A(x) - A(x')\|_1 .$$
The Laplace mechanism adds Laplacian noise to the output of the function $A$ to make it $\epsilon$-local differentially private:
\begin{equation}
\mathcal{A}(x, \epsilon) = A(x) + \Laplace{0}{\frac{\mathrm{Sen}(A)}{\epsilon}}^{d},
\end{equation}
where $\mathrm{Lap}(\mu,b)^{d}$ is a random vector with independent and identically distributed Laplace random variables in each component. 
Here, $A$ is the original forecasting model without randomness, and $\mathcal{A}$ is the differentially private model. $x$ and $x'$ is any possible pair of private data from a single model. 

\subsection{Input-Driven LQR}
\label{sec:InputDrivenLQR}
We model our controller as input-driven LQR motivated by \cite{cheng2021data, li2022adversarial}, which introduce it in terms of data compression and adversarial attacks. 
For clarity, we summarize the derivation here and refer readers to \cite{cheng2021data, li2022adversarial} for details. $\state{}\in \mathbb{R}^{\xtdim}$ and $\action{}\in \mathbb{R}^{\utdim}$ represent the state and action of the controller, and $\Sgt{}\in \mathbb{R}^{\stdim}$ is the external time series.
For time step $\currenttime$, the linear system dynamics are given by:
\begin{equation}
\state{\currenttime+1}=\A\state{\currenttime}+\B\action{\currenttime}+\C\Sgt{\currenttime},
\end{equation}
where $\A\in \mathbb{R}^{\xtdim \times \xtdim}$, $\B\in \mathbb{R}^{\xtdim \times \utdim}$, and $\C\in \mathbb{R}^{\xtdim \times \stdim}$ are the parameters describing how the current state $\state{\currenttime}$, action $\action{\currenttime}$, and external timeseries $\Sgt{\currenttime}$ affect the next state $\state{\currenttime+1}$.
The quadratic cost function is defined with positive-definite matrices $\Q$ and $\R$:
\begin{equation}
\begin{aligned}
\ctrlcost(\bm{\action{}};\bm{\Sgt{}},\state{0})=\sum_{\currenttime=0}^{\horizon}\state{\currenttime}^\top \Q\state{\currenttime}+\sum_{\currenttime=0}^{\horizon-1}\action{\currenttime}^\top\R\action{\currenttime}, 
\quad & \Q\succ0, \R\succ0.
\end{aligned}
\end{equation}
The optimal actions are determined by the initial state and the external timeseries observed by the controller:
$$\bm{\action{}}^*(\state{0};\Sgt{})=\arg\min_{\bm{\action{}}}\ctrlcost(\bm{\action{}};\bm{\Sgt{}},\state{0})=-\K^{-1} (\Lone\state{0}+\Ltwo\bm{\Sgt{}}).$$ 
According to \cite{cheng2021data, li2022adversarial}, 
\begin{align*}
\K = \mathrm{Diag}(\R, \horizon) +\sum_{\currenttime=0}^{\horizon-1}\M_\currenttime^\top \Q \M_\currenttime, \\
\Lone = \sum_{\currenttime=0}^{\horizon-1}\M_\currenttime^\top \Q \A^{\currenttime+1}, \; 
\Ltwo = \sum_{\currenttime=0}^{\horizon-1}\M_\currenttime^\top \Q \N_{\currenttime},
\end{align*}
where $\mathrm{Diag}(\R, \horizon)\in \mathbb{R}^{\utdim\horizon\times\utdim\horizon}$ is a block matrix placing $\horizon$ $\R$ matrices on the diagonal. 
Thus, we define actions generated by perfect timeseries as $\bm{\action{}}^*$ and actions generated by observed, possibly noisy, timeseries as $\bm{\actionhat{}}^*$. 
The two actions and their difference are given by:
\begin{equation}
\bm{\action{}}^*=-\K^{-1} (\Lone\state{0}+\Ltwo\bm{\Sgt{}}), \;
\bm{\actionhat{}}^*=-\K^{-1} (\Lone\state{0}+\Ltwo\bm{\Shat{}}).
\label{eq:OptimalActions}
\end{equation}
\begin{equation}
\bm{\actionhat{}}^* - \bm{\action{}}^* 
= -\K^{-1} \Ltwo (\bm{\Shat{}}-\bm{\Sgt{}}).
\label{eq:actiondiff}
\end{equation}
Eq. \ref{eq:actiondiff} shows the error between actions is linear with respect to the difference of timeseries, and the coefficient is determined by the parameters of the system dynamics, initial state $\state{0}$, and the parameters of the cost function $\Q, \R$. 

Now, we define the increase in the control costs due to forecasting errors as the regret $\Delta \ctrlcost$: 
\begin{equation}
\begin{aligned}
\Delta \ctrlcost
&=\ctrlcost(\bm{\actionhat{}}^*; \bm{\Sgt{}},\state{0}) - \ctrlcost(\bm{\action{}}^*;\bm{\Sgt{}},\state{0}) \\
&=(\bm{\actionhat{}}^*- \bm{\action{}}^*)^\top\K(\bm{\actionhat{}}^*-\bm{\action{}}^*) \\
&= (\bm{\Shat{}}-\bm{\Sgt{}})^\top \underbrace{\Ltwo^\top\K\Ltwo}_{\codesignPsi}  (\bm{\Shat{}}-\bm{\Sgt{}}).
\label{eq:contr_cost_weighted_error}
\end{aligned}
\end{equation}
Note that both $\codesignPsi$ and $\K\succ 0$. 
An intuitive way to explain Eq. \ref{eq:contr_cost_weighted_error} is that extra control cost is quadratic in the forecasting error $(\bm{\Shat{}}-\bm{\Sgt{}})$ since the weight of the error $\codesignPsi$ is positive definite.

%% file: section/SystemModel.tex
\label{system_model}
Imagine that a controller of a ride-sharing operator must assign its taxis to $\xtdim$ locations in a city to serve the customers, and the control decisions are made based on the taxi demand forecasts for different locations. 
At time $\currenttime$, the state $\state{\currenttime} \in \mathbb{R}^{\xtdim}$ represents the difference between the number of free taxis and the number of waiting passengers at ${\xtdim}$ locations, so $\state{\currenttime}>0$ means there are idle taxis, and $\state{\currenttime}<0$ means there are passengers waiting for taxis in the queue. 
The action $\action{\currenttime} \in \mathbb{R}^{\utdim}$ represents the number of taxis sent from $\utdim$ locations to $\xtdim$ queues. 
In our experiment, $\xtdim=\utdim$, but they can be different in general. 

Our system dynamics and control costs are the same as in Sec. \ref{sec:InputDrivenLQR}. 
The demand for taxis is a future timeseries $\Sgt{\currenttime} \in \mathbb{R}^{\stdim}$, and the controller can only estimate it by forecasts $\Shat{\src}$ from different sources $\src=1,...,\srcnum$. 
Here, each source $\src$ can be a firm, such as a cellular network operator, or simply a forecasting model which uses past $\window{\src}$ histories to forecast the next $\horizon$ steps of the future timeseries $\Sgt{}$. 
The forecasts are generated by sensitive private data, thus the sources want to protect their data with $\epsilon$-local differentially private Laplace mechanisms. 
On the other hand, imperfect forecasts lead to an increased cost for the controller. Hence, the controller pays incentives $\incentive{\src}$ to all sources to reduce the noise levels of their Laplace mechanisms. 
The goal of the controller is to minimize its regret $\Delta \ctrlcost$ as defined in Eq. \ref{eq:contr_cost_weighted_error} by allocating incentives to reduce noise levels between different sources. 

In our previous example, the controller wants to keep the queue length $\state{\currenttime}$ close to zero by allocating as few taxis $\action{\currenttime}$ as possible.
We assume that the controller knows its state $\state{\currenttime}$ perfectly, and that the private data are independent for each prediction period. Otherwise,  multiple forecasts at different time steps reveal more information about the same private data, which violates the differential privacy property.

%% file: section/Method.tex
In this section, we use Eq. \ref{eq:contr_cost_weighted_error} to formulate the expected regret $\mathbb{E}[\Delta \ctrlcost]$ of the controller due to Laplacian noise and forecasting errors. We then formulate an incentive allocation problem minimizing the expected regret $\mathbb{E}[\Delta \ctrlcost]$.
We denote \textit{full future} control vectors in bold fonts. Specifically, $\bm{ \action{}}=\action{0:\horizon-1}\in\mathbb{R}^{\utdim\horizon}$, $\bm{\Sgt{}}=\Sgt{0:\horizon-1}\in\mathbb{R}^{\stdim\horizon}$, and $\bm{\state{}}=\state{0:\horizon}\in\mathbb{R}^{\utdim(\horizon+1)}$ for a finite time horizon $\horizon$. 

The timeseries $\bm{\Shat{}}$ are forecasted by $\srcnum$ sources, and we define $\bm{\Shat{\src}}$ as the timeseries sent to the controller from source $\src$ and $\bm{\deltaS{}}$ as the overall forecasting error between the true timeseries $\bm{\Sgt{}}$ and the one observed by the controller $\bm{\Shat{}}$. 
Furthermore, forecasting error $\bm{\deltaS{}}$ can be divided into differential private noise $\bm{\deltaSDP{}}$ and prediction error $\bm{\deltaSForecast{}}$. The first one is caused by the fact that the differential privacy mechanism adds Laplacian noise to the original data, and the latter one is caused by the epistemic uncertainty of prediction.
Hence, we define:
\begin{equation}
\bm{\Shat{\src}} := \bm{\Sgt{}} + \underbrace{\bm{\deltaSDP{\src}} + \bm{\deltaSForecast{\src}}}_{\bm{\deltaS{\src}}}, ~ \forall \src=1,...,\srcnum.
\label{eq:noises}
\end{equation}
We assume that all $\bm{\deltaS{\src}}$ are independent, and that the controller knows the distributions of $\bm{\deltaSDP{}}$ and $\bm{\deltaSForecast{\src}}$ but not the exact values. The distributions are given by:
\begin{equation}
\begin{aligned}
& \bm{\deltaSDP{\src}} \sim \Laplace{0}{\frac{\Ssen{\src}}{\eprivacy{\src}}}^{\stdim\horizon} &, ~ \forall \src=1,...,\srcnum, \\
& \bm{\deltaSForecast{\src}} \sim \mathcal{D}(0,\Sigma_{\src}) &, ~ \forall \src=1,...,\srcnum,
\label{eq:distributions}
\end{aligned}
\end{equation}
where $\mathrm{Lap}(\mu,b)^{\stdim\horizon}$ is a random vector with independent and
identically distributed Laplace random variable in each component, $\Ssen{\src}$, $\eprivacy{\src}$, and $\Sigma_{\src}$ are the local sensitivity of the forecast, the privacy budget, and the variance of the forecasting error at source $\src$, respectively.  $\mathcal{D}(0,\Sigma_{\src})$ is any arbitrary distribution of random vectors with zero mean and variance $\Sigma_{\src}$. 
Here, we only assume the distributions of prediction errors to be zero mean distributions. It is a reasonable assumption since zero mean normal distributions are often used for errors estimations, and our assumption includes all zero mean normal distributions. Also, common regression models, such as linear regression models, have zero bias in their output estimations, thus resulting in zero mean distributions of errors. 

In addition, we use the diminishing properties of logistic functions to capture the marginal effect of incentives $\incentive{}$ on privacy budgets $\eprivacy{}$. 
That is,
\begin{equation}
\begin{aligned}
\eprivacy{\src}(\incentive{\src}) = \frac{\maxeprivacy{\src}}{1+e^{-\edecay{\src}(\incentive{\src}-\midincent{\src})}} 
, ~ \forall \src=1,...,\srcnum,
\label{eq:logistic}
\end{aligned}
\end{equation}
where $\maxeprivacy{\src}$ is the maximum acceptable privacy budget, $\edecay{\src}$ is the increasing rate of privacy budget to incentive, and $\midincent{\src}$ is the function's center of symmetry. 
The logistic function ensures that $\eprivacy{\src}$ asymptotically approaches $\maxeprivacy{\src}$ when $\incentive{\src}$ is infinitely large, so that the privacy is always preserved to a certain level with budget $\eprivacy{\src} < \maxeprivacy{\src}$.

Since the error $\bm{\deltaS{}}$ is a random variable, by Eq. \ref{eq:contr_cost_weighted_error}, the regret of the controller $\Delta \ctrlcost$ is also a random variable. Consequently, our goal is to allocate incentives to different sources in order to increase their privacy budget $\eprivacy{}$ and minimize the expected regret $\mathbb{E}[\Delta \ctrlcost]$. 
Note that in Eq. \ref{eq:contr_cost_weighted_error}, the true control cost $\ctrlcost(\bm{\action{}}^*;\bm{\Sgt{}},\state{0})$ is constant, so minimizing the regret is identical to minimizing the control cost. 
First, by Eq. \ref{eq:contr_cost_weighted_error}, \ref{eq:noises}, \ref{eq:distributions}, and \ref{eq:logistic}, the expected regret can be simplified to:
\begin{equation}
\begin{aligned}
& ~ \mathbb{E}[(\bm{\Shat{\src}}-\bm{\Sgt{}})^\top \codesignPsi(\bm{\Shat{\src}}-\bm{\Sgt{}})] \\
& = Tr\left(\codesignPsi\Sigma_{\src}\right)+ \\
& ~~~~~ 2\left(\Ssen{\src}\times\frac{1+e^{-\edecay{\src}(\incentive{\src}-\midincent{\src})}}{\maxeprivacy{\src}} \right)^2Tr(\codesignPsi \indentity{\stdim\horizon}),
\label{eq:expected_src}
\end{aligned}
\end{equation}
where $\indentity{n}\in \mathbb{R}^{n\times n}$ is the identity matrix, and $Tr(\cdot)$ is the trace function. For simplicity, we denote 
\begin{equation}
\LaplaceVar{\src} = 2\left(\Ssen{\src}\times\frac{1+e^{-\edecay{\src}(\incentive{\src}-\midincent{\src})}}{\maxeprivacy{\src}} \right)^2.
\label{eq:LaplaceVar}
\end{equation}
Note that $\LaplaceVar{\src}$ is convex in $\incentive{\src}$ since the square of an exponential function is convex as well. 
The controller can choose an element-wise linear combination of all $\bm{\Shat{\src}}$ as the timeseries used for control $\bm{\Shat{}}$:
\begin{equation}
\begin{aligned}
\bm{\Shat{}} & = \sum_{\src=1}^{\srcnum} \coeff{\src} \odot \bm{\Shat{\src}},\\ 
\mathrm{s.t.} & \sum_{\src=1}^{\srcnum} \coeff{\src} = \mathbf{1}, ~ 
\coeff{\src} \in \mathbb{R}_{+}^{\stdim\horizon}, \forall \src=1,...,\srcnum,
\label{eq:comb_src}
\end{aligned}
\end{equation}
where $\odot$ is the element-wise product, and $\mathbf{1}$ is the vector of all-ones. 
Therefore, by Eq. \ref{eq:expected_src}, \ref{eq:LaplaceVar}, and \ref{eq:comb_src}:
\begin{equation}
\mathbb{E}[\Delta\ctrlcost] 
= \sum_{\src=1}^{\srcnum} \coeff{\src}^\top \left[\codesignPsi \odot  \left(\Sigma_{\src} + \LaplaceVar{\src} \indentity{\stdim\horizon} \right) \right] \coeff{\src}.
\label{eq:expected_all}
\end{equation}
By the Schur product theorem (\cite[Theorem 7.5.3]{MatrixAnalysis}), the element-wise product of two positive semidefinite matrices is also positive semidefinite, so $\mathbb{E}[\Delta\ctrlcost]$ is convex in all $\coeff{\src}$.
Lastly, the controller allocates incentives to each source to minimize the expected regret $\mathbb{E}[\Delta\ctrlcost]$: 
\begin{subequations}
\label{eq:mini_expected}
\begin{align}
\min_{\substack{\coeff{1},...,\coeff{\srcnum}\\ \incentive{1},...,\incentive{\srcnum}}} & \quad \sum_{\src=1}^{\srcnum} \coeff{\src}^\top \left[\codesignPsi \odot \left(\Sigma_{\src} + \LaplaceVar{\src} \indentity{\stdim\horizon} \right) \right] \coeff{\src} \label{eq:mini_expected_obj}\\
\mathrm{s.t.} 
& \quad \sum_{\src=1}^{\srcnum} \coeff{\src} = \mathbf{1} \label{eq:mini_expected_constrC} \\ 
& \quad \sum_{\src=1}^{\srcnum} \incentive{\src} = \incentive{} \label{eq:mini_expected_constrrho} \\ 
& \quad \incentive{\src} \geq 0, ~\coeff{\src} \geq 0, ~ \forall \src=1,...,\srcnum, 
\label{eq:mini_expected_nonneg}
\end{align}
\end{subequations}
where $\incentive{}$ is the total incentive that the controller can pay to all sources. 
Constraint \ref{eq:mini_expected_constrC} ensures coefficients of all sources 
sum to $1$ (see Eq. \ref{eq:comb_src}), and constraint \ref{eq:mini_expected_constrrho} ensures that the controller can only pay a total incentive of $\incentive{}$ to all sources. The last constraint \ref{eq:mini_expected_nonneg} ensures that all coefficients $\coeff{\src}$ and incentives $\incentive{\src}$ are nonnegative.

Eq. \ref{eq:mini_expected_obj} is convex in $\coeff{\src}$ when $\incentive{\src}$ is fixed and convex in $\incentive{\src}$ when $\coeff{\src}$ is fixed. This property is called \textit{biconvex}, describing a function is convex in two sets of variables independently while the other set is fixed. However, biconvex functions are not convex in both sets of variables simultaneously. 
So far, no algorithm has been found to obtain global solutions of biconvex problems, but it is easy to obtain a local one. 
\cite{Biconvex07} proposed a heuristic – Alternate Convex Search (ACS) to obtain local solutions and showed that if a biconvex function is bounded below in the feasible set, ACS will converge (\cite[Theorem 4.5]{Biconvex07}). It is exactly the case here since we know the expected regret is quadratic. Therefore,  for all $\bm{\Shat{}}$: 
$$\mathbb{E}[\Delta\ctrlcost] = \mathbb{E}[(\bm{\Shat{}}-\bm{\Sgt{}})^\top \codesignPsi(\bm{\Shat{}}-\bm{\Sgt{}})]\geq 0,$$ 
and the feasible set is a subset of all $\bm{\Shat{}}$.

\input{figure_latex/ACS}

The inputs of Alg. \ref{alg:ACS} are a convergence criterion $\eta$ and any feasible point of incentive $\incentive{\src}$ and coefficient $\coeff{\src}$. 
The criterion $\eta$ determines when to terminate, and the feasible point gives the algorithm a point to start with. 
It returns a local solution of Eq. \ref{eq:mini_expected} with the corresponding value. 
In lines \ref{ln:first_loop_starts} to \ref{ln:first_loop_ends}, the algorithm first calculates the initial expected regret. 
Then, the while loop at line \ref{ln:while_loop_start} iteratively searches for a local optimal solution. It first optimizes incentives $\incentive{\src}$ with fixed coefficients $\coeff{\src}$ at line \ref{ln:min_rho} and then does the opposite at line \ref{ln:min_c}. Lastly, the algorithm calculates the new expected regret at line \ref{ln:track_regret} and sees if the difference of the $2$ newest expected regrets is larger than criterion $\eta$. 
In practice, we can modify it to be a running average over the last $M$ measurements.
If it is, the algorithm loops again; otherwise, the algorithm terminates and returns the newest solution and the expected regret. 

\textbf{Limitations:} Our system model assumes that the training set and the testing set of forecasting models have similar distributions so that the covariances of forecasting errors are similar. 
The deviation of the testing distribution from the training distribution is called  \textit{concept drift} \cite{OverviewConceptDrift, SurveyConceptDrift, LearningConceptDrift}. 
In our case, we use the covariance of training set to estimate the one of testing set. 
Our following numerical results show that the training and testing convariances are similar, so \acs~performs well. 
However, when the two covariances are very different, that is, if concept drift occurs, the performance of \acs~will degrade. 

%% file: figure_latex/ACS.tex
\setlength{\textfloatsep}{7pt}
\begin{algorithm}[t]
\caption{Alternate Convex Search (ACS)}\label{alg:ACS}
\renewcommand{\algorithmicrequire}{\textbf{Input:}}
\renewcommand{\algorithmicensure}{\textbf{Output:}}
\begin{algorithmic}[1]
\Require Arbitrary feasible point $(\coeff{1}^0,\incentive{1}^0,...,\coeff{\srcnum}^0,\incentive{\srcnum}^0)$, and convergence criterion $\eta$
\Ensure Local optimal solution $(\coeff{1}^*,\incentive{1}^*,...,\coeff{\srcnum}^*,\incentive{\srcnum}^*)$ and the expected regret $\mathbb{E}[\Delta\ctrlcost]^{*}$

\State $i \gets 0$ \label{ln:first_loop_starts}
\State $\mathbb{E}[\Delta\ctrlcost]^{0} \gets \text{evaluate Eq.\ref{eq:mini_expected_obj} on } (\coeff{1}^0,\incentive{1}^0,...,\coeff{\srcnum}^0,\incentive{\srcnum}^0)$
\State $\mathbb{E}[\Delta\ctrlcost]^{-1} \gets 0$  \label{ln:first_loop_ends}

\While{$\|\mathbb{E}[\Delta\ctrlcost]^{i}-\mathbb{E}[\Delta\ctrlcost]^{i-1}\| \geq \eta$}  \label{ln:while_loop_start}
    \State $\incentive{\src}^{i+1} \gets \arg\min_{ \incentive{\src}} \text{Eq.\ref{eq:mini_expected} for fixed } (\coeff{1}^i,...,\coeff{\srcnum}^i)$  \label{ln:min_rho}
    \State $\coeff{\src}^{i+1} \gets \arg\min_{ \coeff{\src}} \text{Eq.\ref{eq:mini_expected} for fixed } \label{ln:min_c} (\incentive{1}^{i+1},...,\incentive{\srcnum}^{i+1})$
    \State $\mathbb{E}[\Delta\ctrlcost]^{i+1} \gets \text{evaluate Eq.\ref{eq:mini_expected_obj} on } \newline~~~~~~~~~~~~~~~~~~~~~~~~~~~~~~~~~~~~~~(\coeff{1}^{i+1},\incentive{1}^{i+1},...,\coeff{\srcnum}^{i+1},\incentive{\srcnum}^{i+1})$ \label{ln:track_regret}
    \State $i \gets i+1$

\EndWhile
\State \Return $ (\coeff{1}^{i},\incentive{1}^{i},...,\coeff{\srcnum}^{i},\incentive{\srcnum}^{i}) \text{ and } \mathbb{E}[\Delta\ctrlcost]^{i}$
\end{algorithmic}
\end{algorithm}

%% file: section/Experiments.tex
We implemented \acs, the method described in Alg. \ref{alg:ACS}, to show that our theoretical analysis can effectively reduce the regret $\Delta\ctrlcost$. 
We compared our method to a naive heuristic, which uniformly allocates coefficient $\coeff{\src}$ and incentives $\incentive{\src}$ among all sources.
We denote this method as \uniform. 
For \uniform, $\coeff{\src}=\mathbf{1}/{\srcnum}$ and $\incentive{\src}={\incentive{}}/{\srcnum}$ for all sources $\src$.
The initial input of $\coeff{\src}^0$ and $\incentive{\src}^0$ in Alg. \ref{alg:ACS} are the same as \uniform~ for all sources $\src$. 

\input{figure_latex/uber_pickup_distribution}

We evaluated our methods on two timeseries datasets. The first one is a synthetic \textbf{Autoregressive Integrated Moving Average} (ARIMA) timeseries \cite{Harvey1990ARIMA} with parameters $(p,d,q)=(0,1,1)$. The other data set is \textbf{Uber Pickups in New York City} \cite{UberKaggle}, which describes the locations and time of Uber pickups in New York City. The control task of the Uber pickups is described in Sec. \ref{system_model}, while we used Uber data to represent the demand of taxis.
We used additional cost parameters $\Q$ to penalize superfluous waiting queues or insufficient taxi supply at different locations, and used $\R$ to represent the cost of sending taxis to different locations.
We used the timeseries from April to July as the training set and August to September as the testing set  for the forecasting models. We also discretized New York City into $4$ regions and calculate their hourly pickup counts. See Fig \ref{fig:uber_pickup} for the original spatial distribution of Uber pickups in New York City.
We scaled all data to range($0$,$1$) and then fit our forecasting models, so all sensitivities $\Ssen{\src}$ are $1$. 
We used a timeseries forcasting package, \textit{tsai} \cite{tsai}, to train neural network models for the forecasting task.
We now describe the parameters of our control task. 
For ARIMA, $\horizon=10$, $\state{0}=1$, $\xtdim=\utdim=\stdim=1$, $\A=\B=\Q=\R=1$, and $\C=-1$.
For the Uber dataset, $\horizon=5$, $\state{0}=1$, $\xtdim=\utdim=\stdim=4$, $\A=\B=\Q=\R=\indentity{4}$, and $\C=-\indentity{4}$.
See Table \ref{tab:expsettings} for parameters of sources in both experiments.

\input{figure_latex/exp_table}
\input{figure_latex/exp_plots}

We evaluate the performance of \acs~ and \uniform~ on two metrics: control regret $\Delta \ctrlcost$ and forecasting errors $\|\Sgt{}-\Shat{}\|_2$. 
Control regret $\Delta \ctrlcost$ is quadratic in forecasting errors $\|\Sgt{}-\Shat{}\|_2$ (Eq. \ref{eq:contr_cost_weighted_error}), so they are positively correlated. 
We show the results of the ARIMA dataset in Fig. \ref{fig:sub_arima} and the results of the Uber dataset in Fig. \ref{fig:sub_uber}, respectively. 
For both datasets, we first show that the regret $\Delta \ctrlcost$ decreases as the total incentive $\incentive{}$ increases. 
It is natural because all sources receive more incentive to lower their differential privacy noise levels.
The control regret $\Delta \ctrlcost$ converges to a positive constant larger than $0$ because the prediction errors are fixed and the differential privacy budget $\eprivacy{\src}$ eventually converges to $\maxeprivacy{\src}$. 
The mean of coefficient $\coeff{}$ is shown to emphasize how the controller weights the sources based on their prediction errors and the Laplace noise levels. 
When $\incentive{}$ is small, all Laplace noise levels are similar since all sources start with similar $\eprivacy{}$. 
The controller tends to weight source $1$ more because its forecasts are more accurate. 
When $\incentive{}$ increases, first the controller weights source $3$ more because its  $\edecay{3}$ is the largest, so its privacy budget $\eprivacy{3}$ increases rapidly to $\maxeprivacy{3}$. 
Later, when $\incentive{}$ is even larger, all sources' $\eprivacy{}$ approach $\maxeprivacy{}$, so the controller starts to weight source $3$ less. Source $3$ still has the largest weights since its maximum acceptable privacy budget $\maxeprivacy{3}$ is the largest.
The second row of Fig. \ref{fig:sub_arima} and Fig. \ref{fig:sub_uber} shows the true timeseries (black) and the forecasts by \acs~ (green) and \uniform~ (blue) when total incentive $\incentive{}$ is $1$ and $4$. 
\acs~ is more accurate than \uniform, and when the total incentive is larger $\incentive{}$, the forecasts are also more accurate.
We show the distributions of forecasting errors under different total incentives $\incentive{}$ in the right plot of the $2$nd row to emphasize \acs~ effectively reduces forecasting errors and thus reduces the control regret. 

Our results confirm that \acs~ can effectively reduce control regret due to imperfect forecasts, and it reduces the expected regrets by 2.5 and 2.7 times compared to the other benchmark, \uniform. 
Also, in Fig. \ref{fig:arima_uber}, all coefficients $\coeff{}$ are non-binary, meaning that the controller is combining forecasts from different sources. 
Intuitively, since all sources' prediction errors and Laplacian noise are independent, combining them results in $0$ covariance. Hence, it helps to minimize the regret.

%% file: figure_latex/uber_pickup_distribution.tex
\begin{figure}[t]
 \begin{minipage}[c]{0.28\textwidth}
\includegraphics[width=1.0\textwidth, height=4cm]{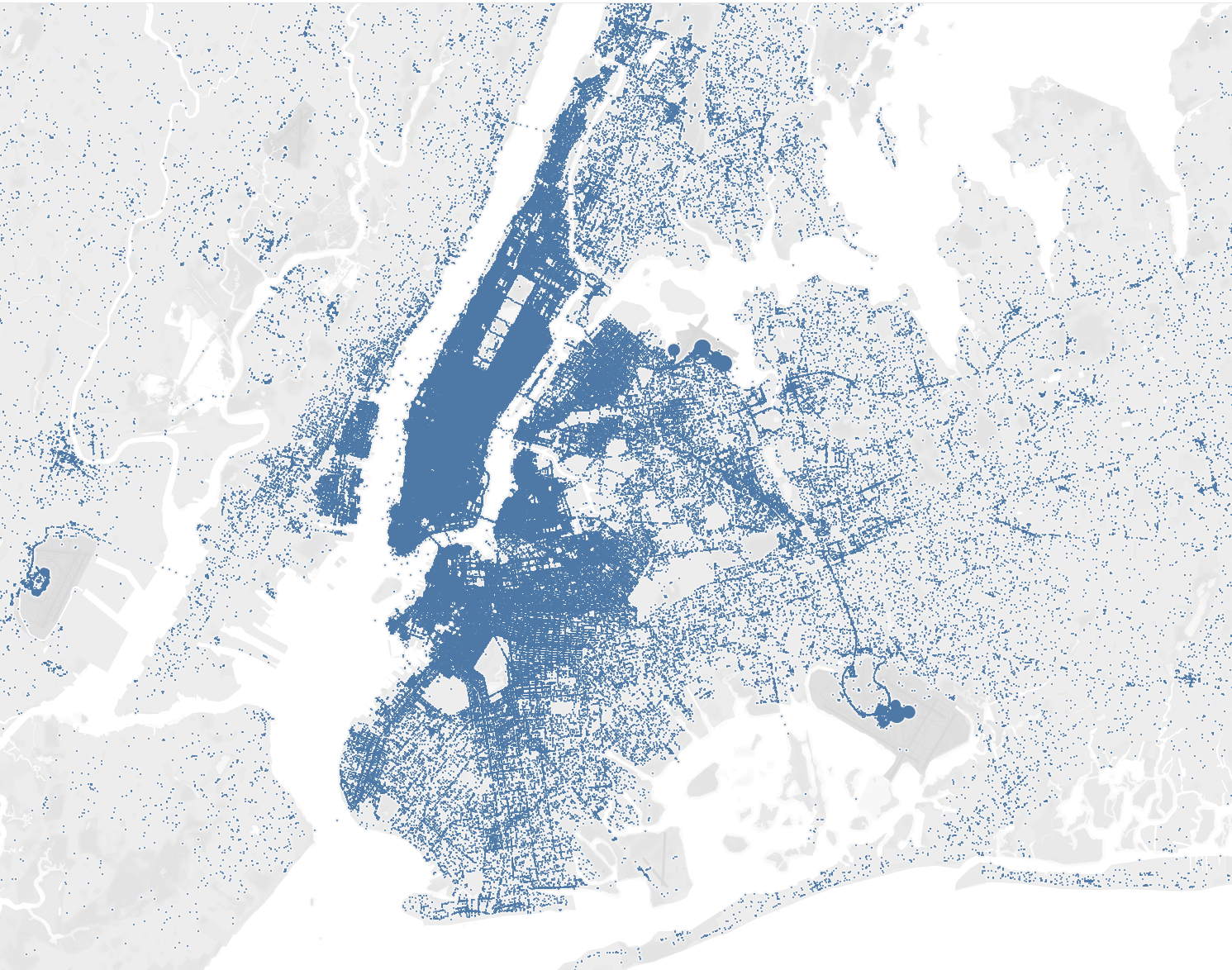}
 \end{minipage}\hfill
 \begin{minipage}[c]{0.2\textwidth}
    \caption{\small{\textbf{Uber Pickups in New York City.}
        The dots in the figure are the pickups in New York City. The more pickups, the larger the dot.
        We discretized the region into $4$ states for the experiments.}}
    \label{fig:uber_pickup}
 \end{minipage}
\end{figure}

%% file: figure_latex/exp_table.tex
\begin{table}[!t]
\begin{tblr}
{
 hlines, vlines,
 rows = {rowsep=1.5pt},
 columns = {colsep=1.5pt},
}
Dataset & $\src$ & $\maxeprivacy{\src}$ & $\edecay{\src}$ & $\midincent{\src}$ & $\window{\src}$ & \makecell{Forecasting Model} \\\hline
\SetCell[r=3]{} ARIMA 
& $1$ & $4$ & $1.5$ & $0.3$ & $10$ & \makecell{Linear Regression} \\\midrule
& $2$ & $8$ & $2.5$ & $0.6$ & $7$ & \makecell{Linear Regression} \\\midrule
& $3$ & $12$ & $3.5$ & $0.9$ & $4$ & \makecell{Linear Regression} \\[2pt]\hline
\SetCell[r=3]{} Uber 
& $1$ & $2$ & $1.5$ & $0.3$ & $10$ & \makecell{TSTransformer \cite{TST}} \\ \midrule
& $2$ & $4$ & $2.5$ & $0.6$ & $7$ & \makecell{InceptionTime\cite{InceptionTime}} \\\midrule
& $3$ & $6$ & $3.5$ & $0.9$ & $5$ & GRU\cite{GRU} \\[2pt] 
\end{tblr}
\caption{\small{\textbf{Detailed Settings of All Sources}: 
For ARIMA and Uber experiments, source $1$ has the most accurate forecast since its window size $\window{1}$ is the largest. In the Uber experiment, source $1$ also has the state-of-the-art timeseries forecasting model. However, its maximum acceptable privacy budget $\maxeprivacy{1}$ is the smallest, meaning it tends to add more Laplacian noise to the forecasts. The converse is true for source $3$, and source $2$ settings are the most moderate.}}
\label{tab:expsettings}
\vspace{-0.5em}
\end{table}

%% file: figure_latex/exp_plots.tex
\begin{figure*}[!]
\centering
\begin{subfigure}[t]{0.75\textwidth}
    \includegraphics[width=\textwidth]{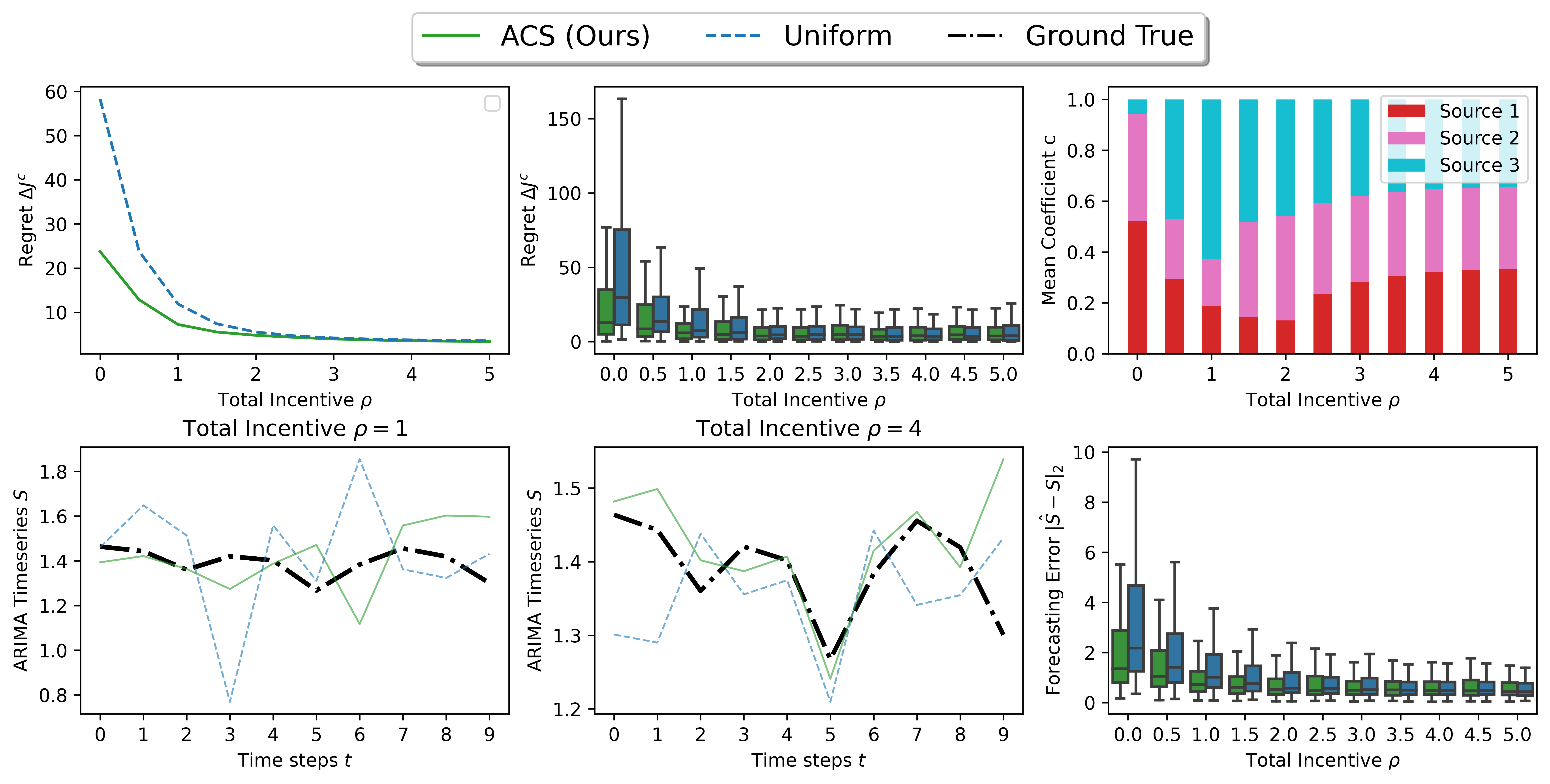}
    \caption{\small{ARIMA timeseries dataset.}}
    \label{fig:sub_arima}
\end{subfigure}
\hfill
\begin{subfigure}[t]{0.75\textwidth}
    \includegraphics[width=\textwidth]{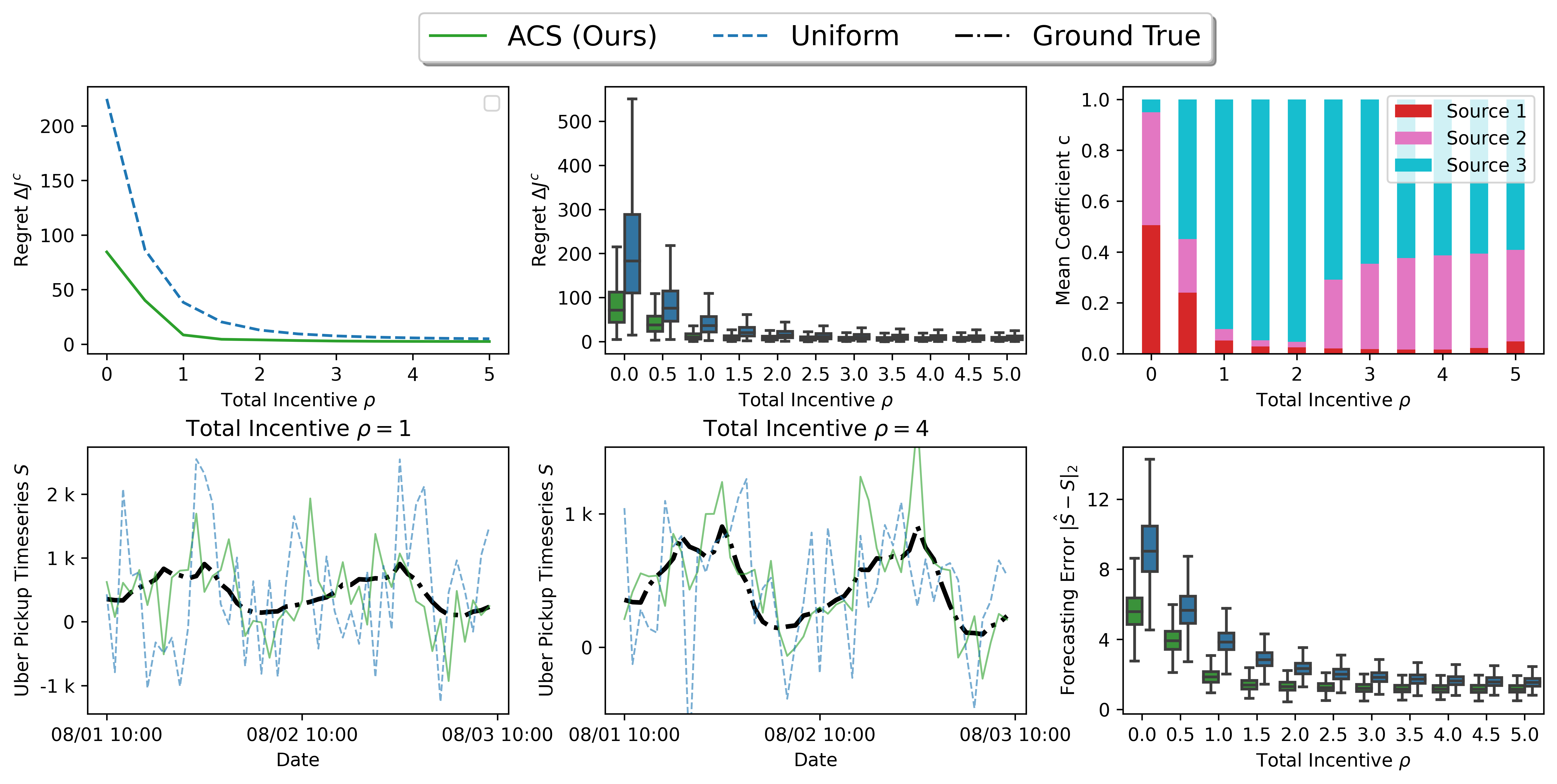}
    \caption{\small{Uber timeseries dataset.}}
    \label{fig:sub_uber}
\end{subfigure}
\caption{
\small\textbf{\acs~outperforms \uniform~in forecasting errors and control regrets}
Fig. \ref{fig:sub_arima} and Fig. \ref{fig:sub_uber} show the results of ARIMA and the Uber demand forecasts.
\textbf{(Row 1)} 
The first $2$ plots on the left in row 1 show the expected control regret (in curves) and actual distributions (in boxplots) of \acs~ (green) and \uniform~ (blue) versus total incentive of the controller $\incentive{}$.
The control regret $\Delta \ctrlcost{}$ decreases as the total incentive $\incentive{}$ increases since incentives reduce the Laplacian noise levels. However, since the trade-off of privacy and incentive is a diminishing function (Eq. \ref{eq:logistic}), the control regret converges asymptotically. 
The right plot shows the mean of coefficient $\coeff{}$ of all sources versus total incentive $\incentive{}$. When $\incentive{}$ is small, the Laplace noise levels are similar among sources so the controller weights the most accurate source more. When $\incentive{}$ is bigger, the effect of Laplace noise levels is more significant than prediction errors, so the controller tends to choose the source with the largest maximum privacy budget $\maxeprivacy{}$. 
Note that the lines in the boxes represent the medians of the distributions, while the curves of regrets are the expected value.
\textbf{(Row 2)} 
The first $2$ plots show the true timeseries (black) and the forecasts by \acs~ (green) and \uniform~ (blue) for $2$ total incentives $\incentive{}$, $1$ and $4$. \acs~ is more accurate than \uniform, which confirms why the control regret of \acs~ is better. Note that these $2$ plots have different scales in the y-axis. Again, the larger $\incentive{}$, the more accurate the forecasts. 
We show the distributions of forecasting errors under different total incentives $\incentive{}$ in the right plot to emphasize our point. We only scaled the timeseries back to the real values when comparing to ground true timeseries; the other plots are calculated with timeseries scaled to range ($0$, $1$). 
The time series of Uber pickups are plotted in $48$-hour increments, so one can clearly see the $2$ cycles in the plots.
}
\label{fig:arima_uber}
\vspace{-1.7em}
\end{figure*}

%% file: section/Conclusion.tex
In this paper, we use the intrinsic prediction errors and privacy preferences of different sources of forecasts to formulate a timeseries forecast selection problem for LQR. 
We use the diminishing properties of logistic functions to capture the marginal effect of incentives $\incentive{}$ on privacy budgets $\eprivacy{}$. 
We then obtain the local optimum of the proposed biconvex optimization problem by \textbf{Alternative Convex Search} (\acs). 
We use linear regression models to forecast ARIMA timeseries and state-of-the-art neural network forecasters to forecast real-world timeseries from Uber. 
Numerical results show that \acs~ reduces the expected regrets by 2.5 and 2.7 times compared to the other benchmark, \uniform. 

In this work, we assume that the private feature data are independent for each prediction period, so that multiple forecasts at different time steps do not cause privacy leakage.
\cite{PriPeARL} proposed a way to allocate privacy budgets among different queries at different time steps to discard this assumption. However, the optimal allocation of privacy budgets $\eprivacy{}$ among different time steps for control remains an open problem. 
We also use the Laplace mechanism to protect private feature data of forecasts. Nevertheless, the information of forecasting models (e.g. parameters) is not protected. 
Hence, another open problem is how to protect information about the forecasting models' architectures and parameters.


%% file: section/Acknowledgement.tex
This work was supported in part by the grants ARL ACC-APG-RTP W911NF1920333, ONR N00014-21-1-2502, ARO W911NF-20-1-0140, and  National Science Foundation grant No. 2133481. Any opinions, findings, and conclusions or recommendations expressed in this material are those of the authors and do not necessarily reflect the views of the National Science Foundation.